# When Knots are Plectonemes

Fei Zheng[1,5,6], Antonio Suma[4], Christopher Maffeo[3], Kaikai Chen[5], Mohammed Alawami[1], Jingjie Sha[6], Aleksei Aksimentiev[3], Cristian Micheletti[2*], Ulrich F Keyser[1*]

1. Cavendish Laboratory, University of Cambridge, CB3 0HE, Cambridge, United Kingdom.
2. SISSA - Scuola Internazionale Superiore di Studi Avanzati, I-34136, Trieste, Italy.
3. Department of Physics, University of Illinois at Urbana–Champaign, 61801-3080, Urbana, USA.
4. Dipartimento Interateneo di Fisica, Università degli Studi di Bari and INFN, I-70126, Bari, Italy.
5. School of Nanoscience and Nanoechnology, University of Chinese Academy of Sciences, 101408, Beijing, China.
6. Jiangsu Key Laboratory for Design and Manufacture of Micro-Nano Biomedical Instruments, School of Mechanical Engineering, Southeast University, 211100, Nanjing, China.

**Abstract**

The transport of DNA polymers through nanoscale pores is central to many biological processes, from bacterial gene exchange to viral infection. In single-molecule nanopore sensing, the detection of nucleic acid and protein analytes relies on the passage of a long biopolymer through a nanoscale aperture. Understanding the dynamics of polymer translocation through nanopores, especially the relation between ionic current signal and polymer conformations is thus essential for the successful identification of targets. Here, by analyzing ionic current traces of dsDNA translocation, we reveal that features up to now uniquely associated with knots are instead different structural motifs: plectonemes. By combining experiments and simulations, we demonstrate that such plectonemes form because of the solvent flow that induces rotation of the helical DNA fragment in the nanopore, causing torsion propagation outwards from the pore. Molecular dynamic simulations reveal that plectoneme initialization is dominated by the applied torque while the translocation time and size of the plectonemes depend on the coupling of torque and pulling force, a mechanism that might also be relevant for *in vivo* DNA organization. Experiments with nicked DNA constructs show that the number of plectonemes depends on the rotational constraints of the translocating molecules. Thus, our work introduces plectonemes as essential structural features that must be considered for accurate analysis of polymer transport in the nanopore.

**Introduction**

The rapid analysis of biopolymers like DNA is a major goal in many biosensing applications[1]. With nanopore sensing, the length and three-dimensional shape of a translocating molecule are extracted from the ionic current signal[2]. Topological conformations like knots or other complex structures are increasingly recognized as crucial targets of single-molecule biosensors for monitoring cellular functions[3–9]. DNA sequencing using nanopore sensors[10] has found numerous applications; however, in such setups, translocation is controlled by molecular motors[11–13] that unzip DNA and remove 3D entangled motifs like knots and plectonemic supercoils from the sequenced strand. In contrast, larger solid-state nanopores have revealed the dynamics of double-stranded DNA (dsDNA) polymers as they traverse through nanoscale confinements since the start of the field in the early 2000s[14–17]. Generally, voltage-driven DNA translocation through a nanopore is often simply described as a quasi-one-dimensional movement with the force applied in the direction of the double helix[18–21]. However, dsDNA in the usual B-form has a right-handed helical structure and thus additional rotational forces were recently observed in experiments and simulations[22–24]. Simply put, the ionic current transfers momentum to the water molecules in the nanopore which results in a torque on the dsDNA. Accordingly, the DNA helix starts to rotate. The additional torque should lead to a rotational movement with consequences beyond the usual picture of simple, force-driven electrophoretic translocation.

Electrophoretic-driven translocation is a process that proceeds far from equilibrium. Especially in experiments where the DNA length is much greater than the nanopore diameter and length[25–27], the dynamics of the DNA segments outside the nanopore influence the translocation process[28]. The translocation time is often so fast that the DNA conformation in front of the nanopore cannot relax during translocation. Effects like tension propagation have been studied both in simulations[28–32] and experiments[14], where a direct consequence of the rapid pulling of the molecules through the nanopore is the possibility to trap, and hence study topological structures like knots[19,33], which are statistically inevitable and yet generally detrimental to the cell functioning[34–37]. Knots form stochastically outside the pore where the DNA strand self-entangles[38–40]. When one end of the DNA strand is electrophoretically pulled into the pore, tension propagates along the strand to the outer part, causing the knot to tighten and ultimately facilitating its translocation[33,38,41]. A characteristic current signature of knots includes a secondary transient spike in addition to the current blockade induced by the linear strand[19,33,42]. Typically, the ionic current signal in a nanopore is translated into the number of strands in the sensing volume. For the simplest trefoil knot, three DNA strands pass the nanopore at the same time and hence the total current amplitude of the knotted signal is 3 times that of the dsDNA current level. More complex knots have been observed in experiments but there is no clear consensus on the exact shape of the knots[19,33].

Here, we use the recent discovery that DNA helix structures experience significant torque and hence rotation during out-of-equilibrium nanopore translocation[23] and harness it to assign numerous ionic current traces to plectonemic structures in addition to well-characterized knots. We present careful measurements of DNA translocation as a function of DNA length, voltage, and torsional constraints. We find that the torsion generated by in-pore electroosmotic flow likely propagates along the strand to the outer DNA segments, which in turn twist the strand into plectonemes. A combination of experiments and simulations indicate that plectonemes can be pulled into the nanopore with three or more dsDNA strands moving inwards concurrently. Our molecular dynamics simulations reveal intricate physical features of plectoneme dynamics during nanopore translocation. Furthermore, we designed nicked DNA structures to show the crucial role of torsional constraints for the formation of plectonemes in the DNA double helix.

**Knots or plectonemes or both in nanopores**

A schematic of the experiment is shown in Figure 1a. A dsDNA molecule (red) is pulled through a conical nanopore by the applied positive potential. The nanopore is large enough (~14 nm in diameter) to allow for the passage of more than one DNA strand. Three possible conformations with three DNA strands side by side are shown in the sketch, the supercoil, the figure-eight knot, and the 'S' shape fold. All these three structures lead to the same ionic current signal as shown in the example trace in Figure 1b. In this signal, we indicate four levels: the baseline '0', the double-stranded DNA level '1', the folded DNA level '2' (not shown in this example) and finally the triple strand level '3' that could represent any of the three structures in Figure 1a. We obtain these current levels from the distributions of an all-current-point histogram (Supplementary Figure 3-6). As the knot is only one of the explanations for the level '3' signal, we call the events of all those possible conformations 'tangled'. In the following, we define the 'tangling probability' as the number of tangled events divided by the total number of nanopore events.

We measured the tangling probability in single nanopore for DNA of 2, 5, 10, 15, 20 and 48.5 kbp in length (blue diamond, Figure 1c). Following previous results[33], we calculated the tangling probability initially attributing all '≥3' events to knots. The tangling probability increases with DNA length which is expected as longer molecules should exhibit more knots. For 48.5 kbp long lambda DNA, we obtained a tangling probability of 62.6%. That is, approximately two out of three translocation events are tangled. The two red lines in Figure 1c show the equilibrium knotting probabilities of DNA molecules at given lengths calculated using Monte Carlo simulations (Supplementary Note 6). The data are for two sets of DNA structural parameters compatible with the experiments, including cross-sectional diameter, d, and persistence length, Lp (see Supplementary Figure 7 for additional conditions). For reference, the simulated knotting probability of lambda DNA is 25.7% and 25.3% for Lp of 40 nm and 50 nm, respectively.

Figure 1c shows a substantial difference between the measured tangling probability (blue symbols) and the calculated maximum knotting probability, a difference that becomes even larger as the DNA length increases. For 48.5 kbp DNA, the tangling probability is more than double the knotting probability, with the absolute difference running up to 37%. Note that the calculated equilibrium knotting probability provides an upper bound for the number of knots detectable via a nanopore translocation measurement[43], as such detectable knots can only originate from self-entanglement of the DNA before the translocation measurement and with no additional knots forming during the translocation process[19,33,44]. The actual experimental knotting probability is expected to be lower than the reference equilibrium value because the knots can slide along and off the DNA during the nanopore translocation without even producing the expected ionic current signature[33,38,41,45]. Hence, we can conclude that the tangled level '≥3' cannot only be knots and must be either the 'S' shape or the plectonemes, Figure 1a.

Next, we examined how the tangling probability (events with levels ≥3) in the longest DNA (48.5 kbp) changes with the applied voltage. Figure 1d shows that the tangling probability increases with the voltage well above the maximum equilibrium knotting probability of lambda DNA (~25%, red line) and that the rate of the increase is consistent among the measurements conducted using different nanopores (blue symbols). That is, the experimental tangling probability is in the range of the maximum knotting probability only at 400 mV and increases from ~35% to more than 60% in the 500 to 800 mV range. This voltage-dependent behavior strongly suggests that the number of tangled events cannot be fully accounted by the expected knotting probability as higher pulling forces cannot create additional knots. Furthermore, the contribution of 'S' structures can be also ruled out as faster translocations are expected to readily unfold the 'S' structures by the propagated tension[33]. Hence, we hypothesize that the excess tangling (≥3 event) probability originates from the formation of plectonemes.

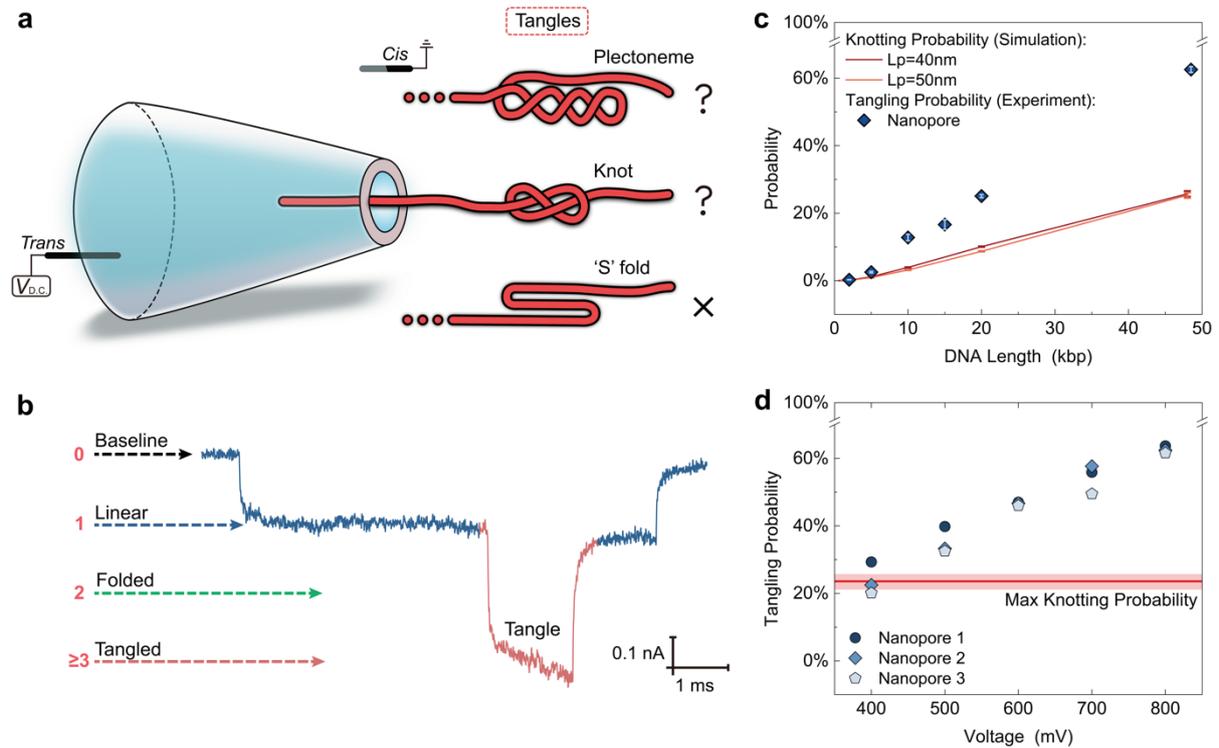

**Figure 1 | Tangling probability is higher than the theoretical knotting probability. a**, Schematic of three tangled conformations, including the plectoneme, the knot, and the 'S' shape, moving through a nanopore. **b**, Ionic current signal from typical tangled events categorized according to four current levels: '0' — current baseline; '1' — linear (single dsDNA strand) translocation; '2' — folded (two dsDNA strands) translocation; '≥3' — Tangled (equal to or more than three dsDNA strands) translocation. The ionic current trace is from a 48.5 kbp lambda DNA translocating through a 14 ± 3 nm (mean ± standard deviation (s.d.)) diameter pore under an applied voltage of +800 mV. The cone semi-angle of this nanopore is 0.05 ± 0.01 radians (mean ± s.d.) based on previous characterization[15]. **c**, Comparison of the experimentally measured tangling probability ('≥3' events) and the equilibrium knotting probability calculated using Monte Carlo simulations. Each diamond symbol shows the experimentally measured tangling probability averaged over three nanopores for the DNA ladder (2, 5, 10, 15, 20, 48.5 kbp). The two lines indicate the simulated knotting probability for the DNA ladder (2, 5, 10, 15, 20, 48 kbp), differing by the assumed persistence length, $L_p$, value: 40 nm or 50 nm. The enlarged views of the 2 and 5 kbp values are provided in Supplementary Figure 7. **d**, Tangling probabilities (lambda DNA) as a function of applied voltage measured in three same-sized nanopores. The horizontal red line is the maximum knotting probability calculated from Monte Carlo simulations for a broad range of $L_p$ and d values.

## Plectonemes not only knots

Plectonemes refer to a twisted DNA structure in which dsDNA strand coils onto itself into a loop-like shape. In the nanopore field, the possibility of plectoneme formation has been so far overlooked as DNA rotation was usually ignored. Recently, the ionic flow in nanopores was shown to empower sustained rotation of DNA helix structures at torques of approximately 0.5 pN·nm per bp at 100 mV·nm$^{-1}$, which is close in magnitude to the torque generated by RNA polymerase to produce supercoils[23,46]. Our conical nanopore are immersed in highly alkaline solution (pH=9) and the electroosmotic flow is significantly enhanced (flow rate ~50 μm$^3$·s$^{-1}$)[47,48]. Thus, we believe the electroosmotic flow in our conical nanopore is sufficient to generate a torque that twists the dsDNA strand and leads to the formation of plectonemes during the translocation process.

The sketches in Figure 2a-c illustrate how a plectoneme initiates and then translocates. The axial electroosmotic flow redirects when it interacts with the DNA helix, imparting a tangential force that produces a torque on the DNA strand (Figure 2c). As the translocation proceeds much faster than relaxation of DNA configuration, the torque subsequently propagates outwards from the nanopore along the strand, which in turn forces the outside segment to rotate about its helical axis. As shown in Figure 2a, the fluid friction interacting with the DNA helix leads to a torque that twists the DNA and may lead to the formation of a plectoneme. Meanwhile, the electrophoretic force pulls the rest of the DNA strand into the pore, eventually translocating the plectoneme (Figure 2b).

The successful translocation of plectonemes relies on the cooperation of electrophoretic force and electroosmotic flow, which are linked in the nanopore. Importantly, the torque depends on the applied voltage and hence we expect that there are more plectonemes forming when the voltage is increased. Figure 2d shows a typical nanopore signal of the plectoneme indicated by the extended level '3' in this example event. The translocation time of a plectoneme inside the pore can reach the scale of milliseconds, which is at least one order of magnitude longer than that of a knot, typically < 100 μs both observed in our nanopore and previous studies[33]. Figure 2e shows typical current blockade signatures that we associate with knotted molecules in our nanopore experiments. In contrast, typical events associated with the plectoneme passage have much longer duration, Figure 2f.

We can thus categorize the tangled events as knots or plectonemes based on the difference in their translocation timescales. Figure 2g shows the normalized duration of the tangled event, which is the duration of a tangled event divided by the duration of the entire translocation event.

Knot signatures are short in duration and appear as a fleeting spike because knots are tightened by the propagated tension (Figure 2e). In contrast, the duration of a plectoneme signal is expected to vary according to the accumulated twists and indeed we observe a wide range of single durations for plectonemes, as illustrated by the examples in Figure 2f. Knowing the total length of the molecules, we can roughly convert the even duration into a tangle size. The knots have an estimated average size of ~140 nm whereas plectonemes are much larger, ~2100 nm (Supplementary Figure 8). Higher-order, hence large-sized, prime knots are rare at the considered DNA lengths (Supplementary Figure 17). As noted above, translocating knots are typically small because they are tightened by tension propagating outwards from the nanopore[45]. In contrast, plectonemes can grow to much larger sizes due to the torque being continuously applied during translocation. The DNA twist introduced by the torque is only partially unraveled by the propagating tension, and thus accumulates in the form of extended plectonemes. The experimental knotting probability exhibits a slight decrease with the applied voltage (Supplementary Figure 9) and is always below the simulation-predicted maximum knotting probability, due to knot sliding.

Finally, we investigated the complexity of the plectonemic conformations. Figure 2h shows the scatter distribution of normalized tangle duration versus normalized current blockade for knots and plectonemes. While most knot and plectoneme events distribute at level '3', a few appear at levels '5' and '7'. Some short-duration events are close to level "4", which we attribute to the limited temporal resolution of our nanopore measurement. Consistent with the equilibrium knot spectrum, and in line with previous research[19,33], we see complex knots in the nanopore signals, such as twin knots, high-order knots, and knots on a folded strand (Supplementary Figure 10,12,17). For plectonemes, we also observe complex plectonemic conformations like a knot on a plectoneme and a complex of plectonemes (Supplementary Figure 11&13).

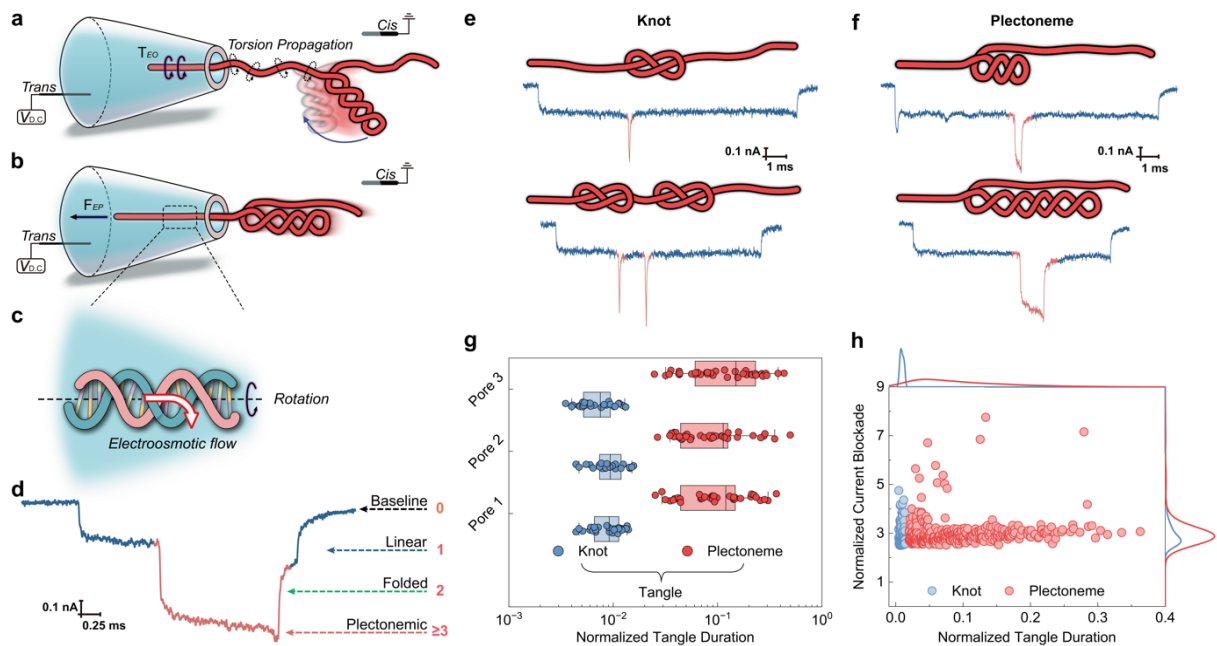

**Figure 2 | Long tangled events are plectonemes. a**, Schematic of the plectoneme formation. A dsDNA strand rotates inside the pore driven by the flow-induced torque. This in-pore torsion propagates to the outside segments of the strand, and forces the strand to coil on itself, thus forming a plectoneme. **b**, Schematic of the plectoneme translocation. Pulled by the electrophoretic force, the plectoneme and another part of the longer molecule are threaded into the nanopore, featuring concurrent translocation of three dsDNA strands. **c**, The origin of the torque. The helical structure deflects the axial electroosmotic flow, the tangential component of which generates the rotation torque. $T_{EO}$ denotes the electroosmosis-generated torque and $F_{EP}$ represents the electrophoretic force. **d**, A representative translocation signal of a plectoneme event. **e**, Examples of knot conformation and corresponding nanopore signals. **f**, Examples of plectoneme conformation and corresponding nanopore signals. **g**, Comparison of the normalized duration of events associated with knots and plectonemes. The statistics are collected from the non-folded events comprising of a knot or a plectoneme measured in three same-sized nanopores. **h**, Scatter plot of normalized current blockade (the level, $I_{\text{tangle}} / I_{\text{linear}}$) versus normalized tangle duration for knots and plectonemes.

**Reproducing plectoneme formation in molecular dynamics simulations**

We performed molecular dynamics simulations of 8 kbp DNA translocation using the validated implicit-solvent model of ref. 49, where DNA is coarse-grained as a twistable elastic discrete chain[50]. The DNA portion in the pore region was simultaneously subject to a longitudinal translocating force and a torque. The pore size and the magnitude of the applied forces were selected consistent with experiments (Figure 3a and Supplementary Note 2). Hundreds of translocation trajectories were collected for different initial equilibrium conformations of the 8 kbp-long strands, with and without knots. The trajectories were analyzed to identify the passage of tangles through the pore (Figure 3b, c); the steric exclusion model[51,52] (SEM, Supplementary Note 8) was used to compute the ionic current blockade from the pore occupation (Figure 3d).

Similar to experiment, the simulated current blockade resulting from a plectoneme translocation through a nanopore was found to exhibit the level '3' signal.

The top panel of Figure 3e shows the percentage of unknotted DNA trajectories presenting at least one tangle event, i.e., ≥3 strands passing through the pore, as a function of the torque and for different translocation forces, starting from initial conformations that were unknotted and torsionally relaxed, hence free of plectonemes. A posteriori analysis and direct inspection of the level '≥3' events revealed the absence of all tangle types in Figure 1a except for plectonemes, confirming that the observed events are associated with the passage of intertwined superstructures formed during translocation. In contrast, knot tangles could be observed only in trajectories where the starting conformation was itself knotted. This is because the translocation process is too short to allow the chain to relax and change its initial topological state. The duration of the knot signals was much shorter than those of plectonemes, again consistent with the experimental observations (Figure 2 and Supplementary Figure 20). Because of their more complex compound entanglement of knots and plectonemes, the simulation traces that originated with having knotted states present more tangle events compared to initially unknotted ones (Supplementary Figure 18).

Using our simulation setup, we could separate the effects of increasing torque and increasing driving force. Inspection of the simulation data reveals that, at a fixed pulling force, various regimes are observed as a function of the torque, Figure 3c. When the latter is sufficiently small, no plectoneme passages are recorded. This is because the DNA twist accumulated on the *cis* side is limited, producing short and loose plectonemes that unravel before reaching the pore entrance. However, as the torque reaches a certain threshold value, the plectonemes formed by the accumulation of twist become sufficiently complex and tightly wound that they can withstand the unravelling action of the propagating tension. These plectonemic structures are preserved as they pass through the pore, yielding the characteristic current drop. Note that the torque threshold for plectoneme passage grows with the pulling force. This is because higher forces are more effective at plectoneme unravelling and thus higher twist rates are needed to observe plectoneme passages. As the torque increases, more plectoneme passages per trajectory are observed (panel 2). A noteworthy feature is that the passing plectonemes become longer as the twist is accumulated more rapidly. This is reflected in the duration of individual plectoneme events growing with applied moderate torque (panel 3). However, because the total pulling force and torque grow extensively with the number of bases inside the pore, the passage of long plectoneme can significantly speed up the translocation of the plectoneme and of the entire chain, too. This effect accounts for the decreasing trends observed at large torque for the entire translocation time and the single plectoneme duration at high force (Figure 3e, bottom panels).

Further parallels between the model and the actual system can be drawn by comparing the simulation data for 8 kbp DNA in Figure 3e with the experimental measurements for lambda DNA in Figure 3f&g. The two key points should be borne in mind for the comparison: (i) the relevant range of torque and pulling forces corresponds to 0.1~0.5 average number of plectonemes, hence in the proximity of the threshold torques, and (ii) unlike in the simulation setup, in experiments the force and torque cannot be varied independently, because they both vary with the applied voltage in an approximately linear fashion[23].

The experimental average number of plectoneme signals increase with voltage (Figure 3f top), while their duration and the entire translocation decrease with voltage (Figure 3f bottom, 3g). All these qualitative features are reproduced by the simulation data. The average number of plectonemes increases both with force and torque and thus also for any of their linear combinations mimicking the effect of voltage increase. Analogous concurrent linear increases of force and torque result in decreasing durations of plectoneme events and of the entire trajectories (Supplementary Figure 19), again in qualitative accord with experiment.

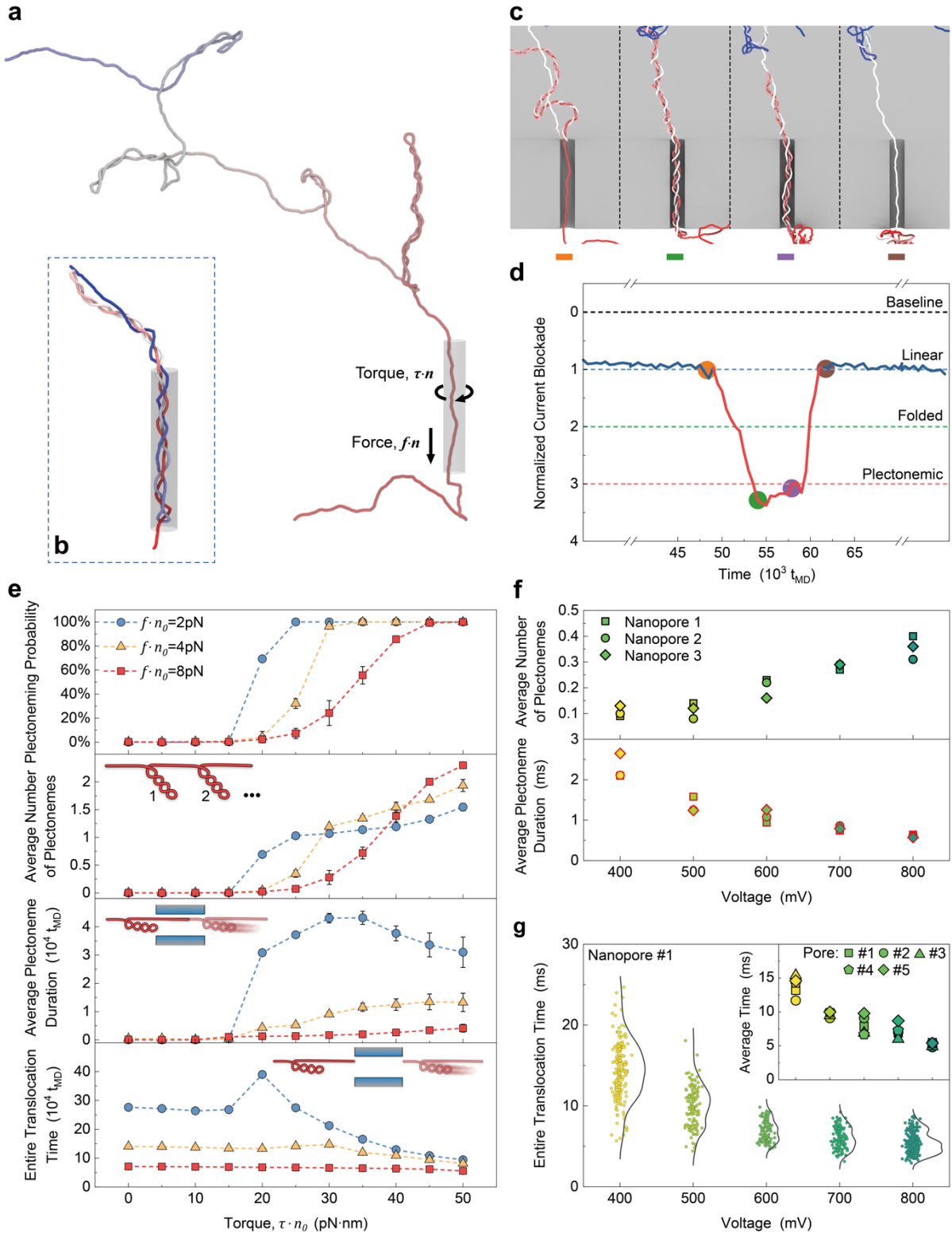

**Figure 3 | Reproduction of plectoneme formation and translocation in molecular dynamics simulations. a**, Snapshot of an 8kbp unknotted DNA chain translocating through a cylindrical pore (shaded cylinder) with a diameter of 14 nm, and a length of 100 nm. $\tau$ and $f$ are, respectively, the torque per base pair and pulling force per base applied to the in-pore part of the DNA molecule. Their values were calibrated to previous electroosmosis-generated torque data in molecular dynamics simulations[23] and optical-tweezer measured electrophoretic forces in conical nanopores[15], respectively
11

(Supplementary Note 7). The total applied pulling force and torque are obtained by multiplying $\tau$ and $f$ by the equivalent number of base pairs in the pore that, $n$. **b**, A snapshot of a plectoneme threading through the nanopore. **c**, Snapshots of four stages of plectoneme translocation through the nanopore, from initial entry to complete exit. **d**, Ionic current signal from plectoneme translocation, with four colored markers highlighting the corresponding stages in **c**. Time is expressed in units of $t_{MD}$, the characteristic simulation time (Supplementary Note 7). **e**, Plectoneme occurrence probability, average number of plectonemes, average duration of single plectoneme and average duration of entire translocation time of all events as a function of the applied torque. The averages are taken over more than 250 independent initial configurations of unknotted and torsionally relaxed chains of 8 kbp length. For reference, the forces and torque reported in the graphs are multiplied by $n_0 = 300$, equivalent to the number of base pairs of a single dsDNA stretch inside the pore. **f**, Average number of plectonemes and average duration of single plectoneme translocation event as a function of applied voltages derived from measurements of lambda DNA in three same-sized nanopores. **g**, Entire translocation time of all events as a function of applied voltages from nanopore experiments. Each scatter point represents an individual event. The lines are half violin distributions which fit the frequency of translocation time values. Inset is the average translocation time obtained from five same-sized nanopores.

**Nicks reduce plectoneme formation**

The simulations suggest that torsional rigidity of DNA is crucial for accumulating plectonemes that can withstand unravelling and pass through the pore. To test this, we designed three different same-length DNA nanostructures having different torsional constraints derived from M13mp18 DNA, including 'intact', 'nicked' and '1nt gap', Figure 4a. The 'intact' one is a dsDNA (7228 bp) cut by restriction enzymes from the double-stranded M13mp18 vector having the phosphodiester bonds chemically linking all adjacent nucleotides. By contrast, the 'nicked' and '1nt gap' DNA have nicking sites evenly distributed along the length of the molecule. We use single-stranded M13mp18 DNA as a scaffold, then hybridize 190 oligonucleotide staples with complementary sequences with it, and thus produce a DNA construct ('nicked') containing 189 nicking sites. Here, each staple contains 38 nucleotides (Figure 4b). The third '1nt gap' construct was created by removing the last nucleotide in each staple resulting in a one-nucleotide gap between every two adjacent staples (37 nucleotides). We note that the three types of DNA molecules are sequence-identical and have the same length. Figure 4c schematically shows how the nicking sites hinder the propagation of torsion along the nanostructures. The prediction is that, without torsion propagation, the rotation is restricted up to the next nicking site, preventing the accumulation of significant twist and the formation of large plectonemes.

For a short DNA molecule like M13mp18 (7.2 kbp) that translocates much faster than lambda DNA (48.5 kbp), it is challenging to discriminate knots and plectonemes according to their duration (Supplementary Figure 15-16). However, because the knotting probability of equilibrated chains depends on large-scale conformations and not on localized defects, the three DNA types are expected a similar number of knots. Thus, the overall tangling probability becomes a probe of how nicking affects plectoneme formation. Figure 4d shows the tangling probability for the three DNA nanostructures, along with another 8 kbp DNA of different sequence used as a control. As expected, the tangling probabilities of these three DNA types follow the order: $P_{\text{intact}} > P_{\text{nicked}} > P_{\text{1nt gap}}$. The control 8 kbp DNA has a slightly higher tangling probability than the other three constructs, which we attribute to the longer length of the former. Furthermore, we examined the voltage dependence of these three DNA types as the formation of plectonemes is significantly affected by the voltage while that of knots is not. Figure 4e shows that the tangling probabilities of all these three DNA types increase with the applied voltage from 400 mV to 800 mV, and, at each voltage, $P_{\text{intact}} > P_{\text{nicked}} > P_{\text{1nt gap}}$. This voltage-dependence further strengthens our conclusion that the torque is propagating along the molecule leading to the accumulation of DNA twist and the formation of plectonemes.

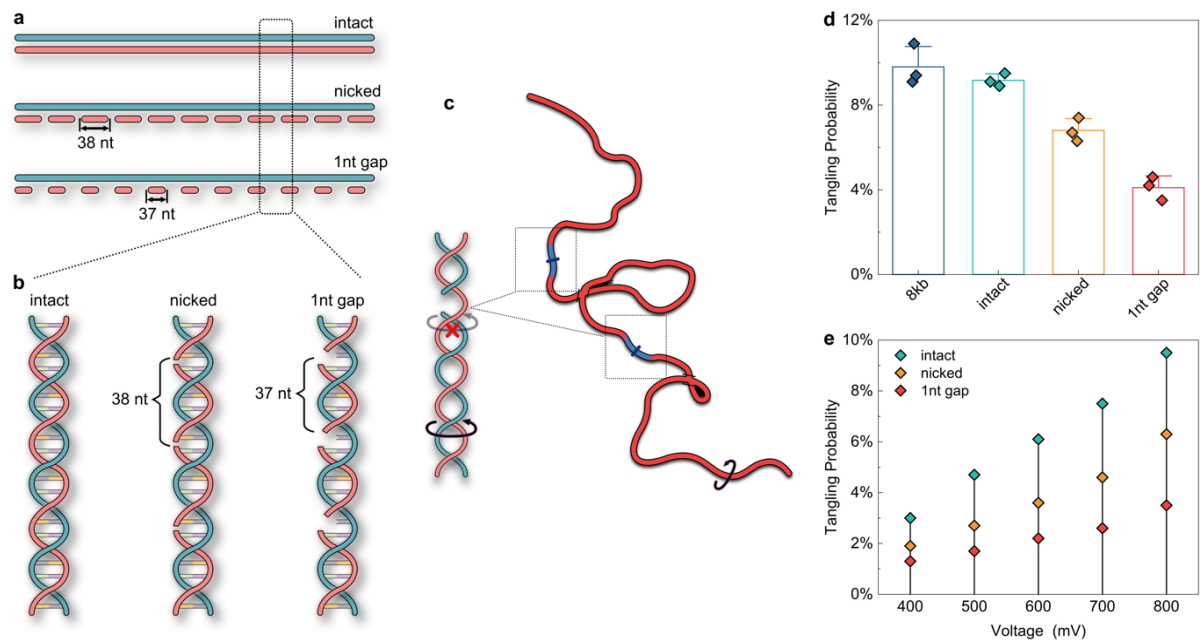

**Figure 4 | Reduction of plectoneme formation resulting from torsion propagation hindrance. a**, Schematic of the 'intact' M13mp18 DNA and the assembly of 'nicked' and '1nt gap' DNA constructs. The 'nicked' and '1nt gap' DNA constructs are synthesized by assembling 190 staples each containing either 38 or 37 nucleotides to the single-stranded M13mp18 scaffold, respectively. **b**, Enlarged view of the helix structures of the three DNA types. The 'nicked' construct has a phosphodiester bond break between each two staples (38 nucleotide distance) whereas the '1nt gap' construct has an additional one-nucleotide gap in each break. **c**, Schematic illustrating how nicking sites hinder the torsion propagation and thus prevent plectoneme formation. **d**, Tangling probabilities of 8 kbp DNA, 'intact' M13mp18 dsDNA, 'nicked' and '1nt gap' DNA constructs measured in three nanopores. **e**, Tangling

probabilities of 'intact' dsDNA, 'nicked' and '1nt gap' DNA constructs as a function of voltage measured in the same pore.

**Concluding remarks**

To conclude, we have demonstrated the presence of torsion propagation on the DNA polymer during its pore-translocation process. The DNA strand outside the nanopore is constantly rotated by this torsion, and thus twisted into plectonemes. We use both experimental nanopore signals and molecular dynamics simulations to evidence that the configured plectonemic structures can survive the unravelling action of the pulling force and translocate through the pore. The formation of plectonemes is strongly dependent on the torque applied on the DNA helix structure, which we show by the increasing overall tangling probability and number of plectonemes per event both in experiments and simulations. Finally, we experimentally employ a nicking strategy to reduce torsion propagation, and thereby hinder the formation of plectonemes.

The implications of our study are numerous. On the one hand, for the purpose of inferring the entangled state of DNA filaments *in vivo*, our results give a first proof of concept demonstration that the same nanopore setup can be used to detect concurrently plectonemes and knots and tell them apart. We recall that knots and plectonemes inevitably arise during the incessant *in vivo* DNA transactions effected by topoisomerase enzymes, which include strand passages, DNA supercoiling and unzipping. Heretofore, these complex forms of DNA entanglement have been detected with electrophoretic techniques, which have been perfected to the point that *in vivo* supercoiling and knotting of circular eukaryotic chromosomes can be measured[53]. However, the complex interplay of the two types of entanglement on electrophoretic mobility, requires distinct measurements for supercoiled and knotted states and an irreversible intervening DNA manipulation: in fact, knots can be reliably detected only after DNA torsion is fully relaxed by nicking. Our results indicate this limitation could be overcome by resorting to nanopore-based setups, thanks to the different duration of the current signals of knots and plectonemes. Obtaining the joint probability distribution of knots and plectonemes *in vivo* would take the characterization of topoisomerase actions to an entirely new level, with significant implications for advancing the current understanding of these enzymes.

On the other hand, for the purpose of biomolecule sensing, the conformational motifs such as folds, knots and plectonemes are not desired as they create complex signals that overlap the information of biopolymers. For instance, knots and plectonemes generate similar current drops as the encoded DNA nanostructures[25] on the polymer scaffold, and thus reduce the detection efficiency. In both cases, our work suggests ways to either avoid introducing spurious plectonemic signals or maintain a certain torque threshold to detect them.

It is an interesting question if minimization of the electroosmotic flow will prevent torque and hence formation of plectonemes or rather reduces their detection as they do not reach the nanopores. In any case, the realization of the importance of torque will allow to tune the solution properties and voltage to optimize the measurements for the respective question. Moreover, as the formation of plectonemes relies on the dynamics of DNA segments outside the pore, accelerating the polymer translocation speed inside the nanopore can likely reduce the plectoneme size as well as their occurrence. Also, given that we have proven that the nicking strategy significantly inhibits plectoneme formation, nicks can be introduced to the DNA polymer as an alternative method.

Vice versa, we have demonstrated that tuning the solution properties or the charge density of the nanopore surface can induce a supercoiled state, and with sufficient torque this state can be maintained and detected as the DNA is pulled into the widening conical pore. Our system represents a viable proxy for studying the extrusion of supercoiled DNA[6].

From a physics-based perspective, our work reveals that the formation of plectonemes depends on several conditions. First is the torque which originates from the interaction between the electroosmotic flow and the DNA helix. This helix structure holds significance in the rotation behavior and subsequent dynamics of the DNA strand outside the nanopore. Apart from the right-handed helix (B-form) used here, left-handed helix DNA was reported to rotate in an opposite direction in simulations[23] while A-form RNA, experiences slower rotation than B-form DNA due to its different helix shape[23]. These rotational differences should have profound influence on the formation of plectonemes. Second, translocation timescale and polymer length are crucial for the occurrence of the plectoneme and its size. Longer DNA polymers are more likely to form large-sized plectonemes as their out-of-pore segments have more time to evolve into plectonemic structures. Additionally, longer polymers feature extended relaxation times, which changes the probability for plectonemes to maintain the structure outside the nanopore.

Both polymer translocation and rotation are out-of-equilibrium processes. From the perspective of energy dissipation during these processes, we show that the torsion propagation outward from the nanopore is another important pathway to dissipate the energy stemming from the applied transmembrane potential. Our work emphasizes that the dynamics of in-pore strand is tightly linked to that of out-of-pore part, by both tension and torsion propagation. Apart from the voltage-driven formation of plectonemes, we expect a salinity gradient across the nanopore could also empower their formation, which is closer to the physiological environment.

**Acknowledgements**

F.Z. and U.F.K. acknowledge funding from an ERC Proof of Concept Grant (PoreDetect 899538) and an ERC consolidator grant (Designerpores 647144). F.Z. acknowledges funding from the China Scholarship Council (202106090221). U.F.K. acknowledges UK Research and Innovation (UKRI) under the UK government's Horizon Europe funding guarantee EP/X023311/1. M.A. was supported by the UK Engineering and Physical Sciences Research Council (EPSRC) grant EP/ S023046/1 for the Sensor CDT. M.A. acknowledges funding from UKSACB scholarship. C.Mi. acknowledges funding from MUR grant PRIN-2022R8YXMR and PNRR grant CN_00000013_CN-HPC, M4C2I1.4, spoke 7, funded by NextGenerationEU. C.Ma. and A.A. acknowledges funding from the Human Frontier Science Project (RGP0047/2020). J.S. acknowledges funding from National Natural Science Foundation of China (52075099, 52361145851).

**Author contributions**

U.F.K., C.Mi. and A.A. conceived the concept of plectoneme formation during nanopore translocation. F.Z. and M.A. conducted the nanopore experiments. K.C. and J.S. set up the nanopore experiment and nanopore fabrication. F.Z. wrote the data analysis scripts with input from M.A. F.Z. and U.F.K. analyzed the experimental data. A.S. and C.Mi. performed DNA analysis of the Monte Carlo and molecular dynamics (MD) simulations. C.Ma. and A.A. performed the ionic current analysis of the MD trajectories. All authors discussed the findings and co-wrote the manuscript.

**Competing interests**

The authors declare no competing interests.